\documentclass[pra,twocolumn]{revtex4-1}
\usepackage[dvips]{graphicx}%
\usepackage{bm,color}
\usepackage{amsmath,amssymb}
\newcommand{\braket}[2]{\langle #1 | #2 \rangle}
\newcommand{\ketbra}[2]{\ket{#1}\!\bra{#2}} 
\newcommand{\ket}[1]{\left |  #1 \right \rangle}
\newcommand{\bra}[1]{ \left \langle #1  \right |}

\begin{document}
\title{Teleportation stretching for single-mode Gaussian channels}

\author{Ryo Namiki}
\affiliation{Institute for Quantum Computing and Department of Physics and Astronomy,
University of Waterloo, Waterloo, Ontario, N2L 3G1, Canada}

\date{\today}
\begin{abstract} 
If a quantum channel is commutable with a certain set of operators, a Choi state of the channel becomes a sufficient resource for any entanglement generation achieved by a single use of the channel with the help of local operation and classical communications (LOCC). This property,   called the teleportation stretchable, could be useful to determine an upper bound of a fundamental rate-loss trade-off for optical quantum key distribution. Unfortunately, the known formulation of  the teleportation-stretching expansion for lossy Gaussian channels  is based on an additional assumption of the Choi state with infinite energy.  In this paper, we present  an entanglement-assisted LOCC protocol with a bounded energy entangled state, and identify a resource state being sufficient for a general entanglement generation protocol through a single use of  lossy Gaussian channels. We also extend this protocol for the cases of  general single-mode Gaussian channels.  Our results provide a regular path to determine an ultimate limit of quantum communication. 
\end{abstract}

\maketitle


\section{Introduction}
Lossy Gaussian channels play a significant role to describe the optical loss and mode mismatches in optical transmission of light fields. 
Quantifying communication capacities for  such channels is a central topic  
of quantum information science 
\cite{HW01,Wolf07,giovannetti2014ultimate}. There are many insightful interplay between entanglement and channel capacities in quantum communication under the local operation and classical communication (LOCC) paradigm \cite{Horo09}. An essential question is what is an ultimate limit of sharable entanglement  per channel use under arbitrary use of LOCC.   Takeoka, Guha, and Wilde introduced the squashed entanglement as an upper bound  on the quantum communication capacity of a quantum channel assisted by unlimited forward and backward classical communication, and established a limit of the fundamental rate-loss trade-off in quantum key distribution   due to  a pure lossy channel 
 \cite{Takeoka2013,Takeoka2014}. 
  Further, a general upper bound of the  rate-loss trade-off on the point-to-point  communication assisted with intermediate stations  has been formulated in Ref.~\cite{Azuma16}. Interestingly, there are observations that the bound is not so far from practically achievable rates \cite{pirandola2015high,xu2015discrete}.

In a recent arXiv paper \cite{PLOB} 
 by Pirandola, Laurenza, Ottaviani, and  Banchi (PLOB), it is claimed that the tight rate-loss trade-off 
 over a pure lossy channel 
with transmission $\eta $ is given by
\begin{align}
-\log_2 (1-\eta ). \label{PLOBbounda}
\end{align}
In order to prove that this is an upper bound,
they introduced a method called the teleportation stretching that essentially implies any entanglement generation through a single use of the channel can be accomplished by an LOCC protocol starting with the Choi state of the channel provided the channel fulfills certain conditions  \cite{Pirandola2015}.  In the case of a finite dimensional systems, the Choi state of a  quantum channel  $\mathcal E $ is normally defined   as 
$ \mathcal I \otimes \mathcal E (\Phi_{\text{MES}})$ with a maximally entangled state $\Phi_{\text{MES}}$.  
As an analogy, PLOB  defined a Choi matrix of the pure lossy channel $\mathcal E_\eta  $ acting on an infinite dimensional system as 
\begin{align}
\rho_{\mathcal E} : = \mathcal I \otimes \mathcal E_\eta  (\Phi_{\text{EPR}}),   \label{Choi}\end{align} where they assumed an ideal Einstein-Podolsky-Rosen (EPR) state $\Phi_{\text{EPR}}$, which can be formally written as  
\begin{align}
 {\Phi_{\text{EPR}}} 
&= \lim_{ \xi  \to 1}  \ketbra{\psi_\xi}{\psi_\xi}.  \label{EPR} \end{align}
Here,   the two-mode squeezed state $\ket{\psi_\xi}$ with $\xi \in (0,1)$ is  a realistic EPR state 
and  given by 
\begin{align}
  \ket{\psi_\xi}: =    \sqrt{1- \xi ^2} \sum_{n=0}^\infty \xi^n \ket{n}\ket{n}. 
  \label{TMSS}  \end{align}
 Due to the teleportation-stretching property, any entanglement generation through a single use of the channel is achieved by an additional LOCC process from the Choi matrix $\rho_{\mathcal E}$. Similarly, any entanglement generation through  an $n$-use of the channel can be accomplished by  an entanglement-assisted LOCC protocol starting from   $\rho_{\mathcal E}^{\otimes n}$. Therefore, any entangled state shared through an $n$-use of the channel with the help of LOCC can be written as 
\begin{align}
\rho_{\textbf{ab}}^n = \bar \Lambda (\rho_{\mathcal E}^{\otimes n}),  \label{LOCC} \end{align}
where $\bar \Lambda$ stands for an LOCC process, and the subscript of  $\rho_\textbf{ab}$ denotes the following condition \cite{PLOB}: The sender Alice prepares an arbitrary quantum state $\rho_{\textbf{a}a_1a_2\cdots a_n}$, and sends  an $n$-mode subsystem $a_1a_2\cdots a_n$ to the receiver Bob   with the $n$-channel use; The final subsystem possessed by Bob is specified by the index $\textbf{b}$.   Given that the final state $\rho_{\textbf{ab} }^n$ is delivered through an LOCC protocol as in Eq.~\eqref{LOCC}, the monotonicity of entanglement implies  that 
 the  amount of entanglement shared between the bipartite system $\textbf{ab}$ is  bounded from above as 
  \begin{align}
E_R (\rho_{\textbf{ab} }^n) \le E_R  (\rho_{\mathcal E}^{\otimes n}),  
\label{eq1} \end{align}
where $E_R $ is the relative entropy of entanglement (REE), but this relation could hold for any entanglement measure. Due to the additivity of $E_R$, we may obtain
 \begin{align}
E_R (\rho_{\textbf{ab} }^n) \le E_R  (\rho_{\mathcal E}^{\otimes n}) \le n E_R  (\rho_{\mathcal E}) . 
\label{eq2} \end{align} From this relation, 
 the amount of entanglement generated per channel use  could be bounded as 
\begin{align}
\frac{E_R (\rho_{\textbf{ab} }^n)}{n}  \le \frac{ E_R  (\rho_{\mathcal E}^{\otimes n}) }{n} \le E_R (\rho_{\mathcal E} ) \le - \log_2(1- \eta ). \end{align}
Notably, the rightmost term $ - \log_2(1- \eta )$  is determined by taking the limit $\xi \to 1$ for $E_R( \mathcal I \otimes \mathcal E_ \eta (\psi_\xi ) )$  \cite{PLOB}. 
Notably, the rightmost term $ - \log_2(1- \eta )$  is determined by taking the limit $\xi \to 1$ for $E_R( \mathcal I \otimes \mathcal E_ \eta (\psi_\xi ) )$  \cite{PLOB}. 
Hence, the RHS of Eq.~\eqref{EPR} reads that the limitation $\xi \to 1$ is taken \textit{eventually}, and until that instance, the ideal EPR state $\Phi_\text{EPR}$ has to be regarded as a normal state since the RHS  of Eq.~\eqref{EPR} is indeterminate by itself. On the other hand, the central relation in Eq.~\eqref{LOCC} due to the teleportation-stretching property, is derived under the assumption of  the ideal EPR state and thus the limit operation $\xi \to 1$ is likely to be conducted before Eq.~\eqref{LOCC}.  To this end, $\rho_{\mathcal E}$ in the RHS of Eq.~\eqref{LOCC} seems already indeterminate, and this could make 
 the subsequent relations [i.e., Eqs.~\eqref{eq1} and ~\eqref{eq2}]  meaningless. This incoherence is obviously  coming from the  assumption of the ideal EPR state in Eq.~\eqref{EPR}, and verifying the mathematical validity in heuristic use of such a fictitious state seems difficult. 

  In this paper, we present a conventional theory of the teleportation-stretching expansion  for Gaussian lossy channels,  and provide the relation essentially equivalent to  Eq.~\eqref{LOCC} without invoking additional  assumptions.  We also extend the result for general single-mode Gaussian channels. Thereby, we address the question in a related work \cite{Namiki13a}: What is a sufficient entangled resource state shared between Alice and Bob for them to prepare an output state $\mathcal E (\rho)$ at Bob's station given an arbitrary input state $\rho$ at Alice's station under the restriction of LOCC.  

This paper is organized as follows.
We start  reviewing the teleportation-stretching property for a finite dimensional system, and develop its counterpart for the pure lossy channel in Sec.~II. A general result for single-mode Gaussian  channels is presented in Sec.~III.   We conclude this paper  in Sec.~IV.

 \section{Teleportation stretching}
\subsection{Finite dimension}
We start  reviewing the teleportation-stretching property for a finite dimensional system associated with the  standard quantum teleportation process for a $d$-dimensional system  \cite{QTeleport}. Let us write a standard basis $\{ \ket{j}\} _{j=0}^{d-1}$.   Let $\ket{ \phi_{0}} = \sum_{j=0}^{d-1}\ket{j}\ket{j}$ be a maximally entangled state, and  
$\{ \ket{\phi_k} = (\openone \otimes \sigma_k )\ket{\phi_0}\}$  with $k =0,1, \cdots d^2 -1$ be a set of Bell states where $\openone $ is the identity operator and $\{\sigma_k\} $  represents  a set of unitary operators. 
We define a teleportation set $\{T_k\}$ associated with a quantum teleportation process from system $A$ to system $B$ using the maximally entangled state ${ \phi_{0}}$ between $A^\prime \! B$  as 
\begin{align}
T_k :=  _{A\! A^\prime}\!\! \braket{\phi_k}{\phi_0}_{A^\prime \! B} = \sigma_k \sum_{j=0}^{d-1} \ket{j}_B\!\bra{j}_A. \label{finiteT}
\end{align} This implies  an arbitrary  input state in system A, say $\rho_A$,  is transferred as 
\begin{align}
T_k  \rho _A T_k^\dagger =  \sigma_k \rho_B \sigma_k  ^\dagger.   
\label{finite}\end{align} Thereby, adjustment of the additional unitary operators $\{\sigma_k\}$ accomplishes the standard  teleportation process. 

A quantum channel  $\mathcal E$  is called teleportation stretchable with regard to  $\{ T_k\}$  \cite{Pirandola2015} if there exists a set of unitary operators $\{ U_k\}$  such that 
\begin{align}
\mathcal E ( T_k \rho  T_k^\dagger ) = 
 U _k   \mathcal E (  \rho )U _k^\dagger   \label{tss}
\end{align}  holds for   any input state $\rho $ and any $k$.
This property can be equivalently expressed as 
\begin{align}
 \mathcal L_{U _k^\dagger} \circ \mathcal E   \circ \mathcal L_{T_k }  ( \rho ) = \mathcal E(   \rho)  \label{t-stretch}
\end{align} when we use a shorthand notation: 
\begin{align}
\mathcal L_U  (\rho ):= U  \rho U^\dagger. 
\end{align}
The main implication of Eq.~\eqref{t-stretch} is that the expression in the LHS can be regarded as an LOCC process based on the shared entanglement $\phi_0$. To be specific, knowing the index of Bell measurement $k$,  the owner of system $B$ can obtain the channel output $\mathcal E (\rho)$ 
 applying a unitary operation locally, as in the final step of the standard quantum teleportation process. 
Therefore, any state transmission through the channel $\cal E$ can be faithfully simulated by an entanglement-assisted LOCC protocol.  
One can simulate the channel action by locally applying the channel $\mathcal E$ after a standard quantum teleportation protocol when the maximally entanglement is shared \cite{Namiki13a}. 
In contrast to this fact, the expression in Eq.~\eqref{t-stretch} tells us that such a  perfect channel simulation can be done  with using a \textit{non-maximally} entangled state corresponding to the Choi state   
\begin{align}
\rho _{\mathcal E }:=\mathcal I  \otimes \mathcal E ( \phi_0), \label{FCHOI}
\end{align}
 whenever  the channel $\mathcal E$ fulfills the commutation property described in Eq.~\eqref{tss}. An example of teleportation stretchable channels is Pauli channels \cite{Pirandola2015}.

Note that the property called the LOCC composability of quantum channels was investigated in Ref.~\cite{Namiki13a}. There, the main question is to identify an entangled state shared between Alice and Bob from which they can prepare an output state $\mathcal E (\rho)$ at Bob's station given an arbitrary input state $\rho$ at Alice's station under the use of LOCC.  The relation in  Eq.~\eqref{t-stretch} shows that the Choi state $\rho _{\mathcal E }$ in Eq.~\eqref{FCHOI} is sufficient to accomplish this task when the channel $\mathcal E$ is the teleportation stretchable.  On the other hand,  it is shown in Ref.~\cite{Namiki13a} that a maximally entangled state whose Schmidt rank is equal  to the Schmidt number of the  Choi state $\rho _{\mathcal E }$ is sufficient for this task.

 \subsection{Infinite dimension with a realistic EPR state}%

In order to consider  the teleportation-stretching property in an infinite dimension, an obstacle is that  the maximally entangled state $\phi_0$ in the limit $d \to \infty$ becomes indeterminate. Instead, we may use the  two-mode squeezed state $\psi_\xi$ in Eq.~\eqref{TMSS} as an effective maximally entangled state as
in the continuous-variable (CV) quantum teleportation scheme by Braunstein and Kimble \cite{Braunstein-Kimble98}. 
As well as   the form of the maximally entangled state, there should be notable differences in the form of Bell states $\{\phi_k \}$. For a two-mode  bosonic field,  we may use 
the simultaneous eigenstates of EPR operators, 
\begin{align}
\ket{\Phi_{0}} 
&:=\frac{1}{\sqrt {2\pi}} \int_{x \in \mathbb{R}} \ket{x}\!\ket{x} dx. 
 \label{CV-Bell} \end{align} where 
  $\ket{x}$    represents the eigenstate of the position quadrature $\hat x $, and is normalized by the Dirac-$\delta$~function as $\braket{x}{x^\prime} = \delta (x-x^\prime)$.  CV-Bell states can be  defined as 
 \begin{align}
\ket{\Phi_\beta} 
:=( \openone \otimes  D^\dagger (\beta ) )\ket{\Phi_0}  
\label{cvBell}
\end{align} where  $ D (\beta ) = e^{\beta a  ^\dagger - \beta ^* a }$ is a displacement unitary operation  with the displacement amount  $\beta= \frac{1}{\sqrt 2} (x+ip)  \in \mathbb{C}$. 
A CV-Bell measurement is associated with the following resolution of the identity 
\begin{align}
\iint\ketbra{\Phi_\beta}{\Phi_\beta}_{A^\prime \! B}dxdp = \openone_{A^\prime \! B}. 
 \end{align}
 Note that   CV-Bell states $\{ \ket{\Phi_{\beta}}\}_{\beta \in \mathbb{C}} $ have the normalization of  the Dirac-$\delta$ function, and should not be regarded as physical state vectors  (see, e.g., Chapter 4 of Ref.~\cite{Peres1994quantum}).

Another essential ingredient for the teleportation-stretching property for lossy Gaussian channels  is  the property of the displacement covariance. 
 For a  pure lossy channel $\mathcal E_\eta $ with the transmission $\eta  \in (0,1)$, it holds that  
\begin{align}
\mathcal E_\eta  ( D(\beta )\rho   D^\dagger(\beta ) ) = 
  D(\sqrt \eta  \beta )  \mathcal E_\eta  (  \rho )  D^\dagger(\sqrt \eta \beta )  .  
\label{cv-t-stretch}\end{align} 
It could be worth noting that  a formulation of the teleportation-stretchable property almost parallel to the finite dimensional case is  possible 
if we assume the CV-Bell state $\Phi_0$ is a physical state (See Appendix~\ref{Note1}).

We consider the stretching process with a two-mode squeezed  
state $\psi_\xi$ defined in Eq.~\eqref{TMSS}.  Let us use the Glauber-Sudarshan $P$ representation for an arbitrary input state as 
\begin{align}
\rho
 = \int_{\alpha \in \mathbb{C}} P_\alpha \ketbra{\alpha}{\alpha} d^2 \! \alpha \label{P-rep}, 
\end{align} where $\ket{ \alpha} = D(\alpha ) \ket{0} $ denotes a coherent state. Suppose that $\rho$ is in system $A$ and $\psi_\xi$ is in system $A ^\prime \! B$, and that the CV-Bell measurement will be  performed on system $A\!  A^\prime $, as in  the finite dimensional case of  Eqs.~\eqref{finiteT}~and~\eqref{finite}.  
 From Eqs.~\eqref{TMSS}, \eqref{cvBell},  \eqref{cv-t-stretch}, and  \eqref{P-rep}, and the property of the displacement operators,  a straightforward calculation leads to   
\begin{align}
&T_{\beta,\xi }\rho T_{\beta,\xi }^\dagger  \nonumber \\
:=&\bra{\Phi_\beta}  (\rho
\otimes \ketbra{\psi_\xi }{\psi_\xi})_{A\!A^\prime\! B} \ket{\Phi_\beta}_{ A\!A^\prime}\nonumber \\
=& \int P_\alpha \ _{ A\!A^\prime}\!\braket{\Phi_\beta}{\alpha}_A  \  (\ketbra{\psi_\xi }{\psi_\xi})_{A^\prime\! B \ A}\!\braket{\alpha}{\Phi_\beta}_{ A\!A^\prime} d^2 \! \alpha \nonumber \\
=& \frac{1}{2 \pi}\int P_\alpha  \bra{\alpha^*}   D_{A^\prime} (\beta )  \ketbra{\psi_\xi }{\psi_\xi}  D_{A^\prime}^\dagger (\beta ) \ket{\alpha^*}_{A^\prime} d^2 \! \alpha \nonumber \\
=& \frac{1}{2 \pi} \int P_\alpha  \braket{\alpha^* +\beta^* }{\psi_\xi }  \braket{\psi_\xi}{\alpha^*+\beta^* } d^2 \! \alpha \nonumber \\
=& \frac{1- \xi^2}{2\pi} \int P_\alpha e^{- (1- \xi^2 )\left| \alpha +\beta  \right|^2 } \ketbra{\xi ( \alpha  +\beta ) } {\xi (\alpha +\beta )} d^2 \! \alpha \nonumber \\
=&  \frac{1- \xi^2}{2\pi} D (\xi \beta ) \left(  \int P_\alpha   e^{- (1- \xi^2 )\left| \alpha +\beta  \right|^2 }   \ketbra{\xi  \alpha }{\xi\alpha}   d^2 \!  \alpha \right )  D^\dagger (\xi \beta ) \nonumber \\
=&  D (\xi  \beta )  \left(  \mathcal E_ {\xi^2} \left( \tilde \rho_ \beta 
 \right ) \right) D^\dagger (\xi \beta ),   \label{body}
 \end{align} where, in the final step,     
  we used the property of the lossy channel,  that is,  $\mathcal E_ {\xi^2}(\ketbra{\alpha}{\alpha})  =  \ketbra{\xi \alpha}{\xi \alpha}$ holds for any coherent states $\ket{\alpha}$ with $\alpha \in \mathbb{C}$,  and  defined 
 \begin{align}
\tilde \rho_ \beta :=   \frac{1- \xi ^2}{2\pi}  \int P_\alpha   e^{- (1- \xi^2)\left| \alpha +\beta  \right|^2 }   \ketbra{ \alpha }{\alpha}   d^2 \!  \alpha.   \label{tilro} \end{align}
For later use we note
 the following relation: 
 \begin{align}
\iint \tilde \rho_ \beta 
{d\!x d\!p } =2 \int \tilde \rho_ \beta 
{d^2\! \beta} =  \int P_\alpha    \ketbra{ \alpha }{\alpha}   d^2 \!  \alpha = \rho . \label{norm-beta}\end{align}

With the help of Eq.~\eqref{cv-t-stretch} and another property of the lossy channel $\mathcal E _\eta \circ \mathcal E_{\eta^\prime } =  \mathcal E_{\eta \eta^\prime }$,  Eq.~\eqref{body} implies 
\begin{align}
   \mathcal E_{\eta} \left( T_{\beta,\xi  }\rho T_{\beta, \xi}^\dagger  \right)
=&   \mathcal E_{\eta} \left(  D ( \xi  \beta )  \left(  \mathcal E_ {\xi^2}\left( \tilde \rho_ \beta  \right ) \right) D^\dagger (\xi \beta )\right)  \nonumber \\
=&    D (\sqrt{  \eta } \xi  \beta )  \left(  \mathcal E_{\eta} \left( \mathcal E_ {\xi^2} \left(\tilde \rho_ \beta  \right ) \right)\right)   D^\dagger (\sqrt {  \eta }\xi \beta )\nonumber \\
=&     D (\sqrt{ \eta } \xi \beta )  \left(  \mathcal E_{\eta\xi^2} \left( \tilde \rho_ \beta  \right)\right)   D^\dagger (\sqrt {  \eta } \xi \beta ). \label{Sub}\end{align}
This relation is slightly different from the formula in Eq.~\eqref{tss}, and  the teleportation-stretching property cannot hold for the realistic CV-teleportation set $\{T_{\beta, \xi}\}$. However, we can show  a modified version of such a property sufficient to simulate the channel action. We may say that the pure  lossy channel is \textit{effectively  teleportation-stretchable} regarding the following relation: 
\begin{align}
 \int   D^\dagger (\sqrt{   \eta } \xi \beta )    \mathcal E_{\eta} \left( T_{\beta,\xi }\rho T_{\beta, \xi}^\dagger  \right)  D (\sqrt{  \eta }\xi \beta )   {d\! x d\! p}  
=        \mathcal E_{\eta\xi^2} \left( \rho  \right).  \label{Main}\end{align}
This relation can be readily obtained by an application  of   $ \mathcal L_{  D^\dagger (\sqrt{  \eta } \xi \beta )} $ and integration  over $\beta$  for both sides of Eq.~\eqref{Sub} with the help of Eq.~\eqref{norm-beta}. 
As  a counterpart of the expression in Eq.~\eqref{t-stretch}, the following representation would be helpful:
\begin{align}
 \int  \left( \mathcal L_{  D^\dagger  (\sqrt{  \eta } \xi \beta )}   \circ \mathcal E_{\eta} \circ \mathcal L_{ T_{\beta, \xi  }} (\rho ) \right )  d\! x d\! p 
=&  \mathcal E_{\eta\xi^2  }  (\rho ) \label{MainR} . \end{align}  
In both the representations in Eqs.~\eqref{Main} and \eqref{MainR},  the expression of the LHSs  is regarded as an LOCC process based on the realistic EPR source $\psi_\xi $ and a single use of the lossy channel $\mathcal E_\eta$. Therefore, the central implication of our result is that any transmission through a lossy channel  $\mathcal E_{\eta}$ can be simulated by an entanglement-assisted LOCC  with the resource state generated through another lossy channel $ \mathcal E_ {\eta\prime }$ and a two-mode squeezed state $\psi_\xi $.  Let us restate this relation as a theorem:

\textbf{Theorem}.---Let $\mathcal E _\eta$ be a pure lossy channel with trasmission $\eta \in (0,1)$. For any two-mode state $\rho_{AB}$, there exists another pure lossy channel $\mathcal E_{\eta ^\prime }$ with  $\eta^\prime \ge \eta $, a two-mode squeezed state $\psi_\xi$, and an LOCC process $\Lambda $, such that
\begin{align}
\mathcal I \otimes \mathcal E_ {\eta } (\rho_{AB}  ) = \Lambda \left (\rho_{\mathcal E ^\prime}  \right),  \label{Theorem}
\end{align}
where an effective  Choi state $\rho_{\mathcal E ^\prime}$ is defined as 
\begin{align}
\rho_{\mathcal E ^\prime}:=  \mathcal I \otimes \mathcal E_ {\eta\prime } (\psi_\xi  ).\end{align}The LOCC process can be represented by the effective teleportation-stretchable condition of Eq.~\eqref{Main} (or equivalently Eq.~\eqref{MainR}). The parameters $(\eta^\prime,\xi )$ can be chosen so as to satisfy  $\eta ^\prime = \eta / \xi^2 $.

\textbf{Proof}.---Obvious from the effective teleportation-stretchable  property in Eq.~\eqref{Main} [or equivalently, Eq.~\eqref{MainR}]. \hfill$\blacksquare$

Physically, the use of the realistic EPR state $\psi_\xi$ implies a teleportation resource of a non-maximally entangled state. Then, the resultant teleportation cannot be perfect, and a part of  the  imperfection is equivalently induced by a lossy channel  as in Eq.~\eqref{body}. Thereby,  
 compensation of the gain seems be essential if one insists that the unit-gain condition is crucial, although the effect off loss is inherently irreversible  \cite{Namiki07,Namiki-Azuma13x}.
  In contrast, our goal here is to simulate the lossy channel $\mathcal E_\eta$, and we thus rather make use of the imperfection as a natural tool to execute the  precise simulation.

  \subsection{Step for  a regular proof} 
  Now we can establish a regular step to find the bound of Eq.~\eqref{PLOBbounda}  claimed in Ref.~\cite{PLOB} as follows. 
Since, any two-mode state transmission can be simulated by an entanglement-assisted LOCC protocol as in the RHS of Eq.~\eqref{Theorem}, any state transmission plus LOCC operation can be also simulated by an entanglement-assisted LOCC protocol from the Choi state $ \rho_{\mathcal E ^\prime}$.  We can apply  this transmission-plus-LOCC process for entanglement generation based on the transmission of an arbitrary $n$ modes. 
Therefore,
any entanglement generated by an $n$-use of channel and LOCC can be achieved by an LOCC protocol when  $ \rho_{\mathcal E ^\prime}^{\otimes n}$ is shared between the bipartite system $\textbf{ab}$. This establishes a finite-energy counterpart of Eq.~\eqref{LOCC}  as 
\begin{align}
\rho_{\textbf{ab}}^n = \bar \Lambda ( \rho_{\mathcal E ^\prime}^{\otimes n})  
. \label{LOCCR} \end{align} 
 Hence, we can find an upper bound of entanglement due to the LOCC monotonicity without additional assumptions. For REE, it holds that  
  \begin{align}
E_R (\rho_{\textbf{ab} }^n) \le E_R  (\rho_{\mathcal E^\prime}^{\otimes n}) . 
\label{eq1R} \end{align}
We can repeat the flow of Eqs.~\eqref{eq1}~and~\eqref{eq2}, and eventually take the limit $\xi \to 1$ so as to reach the upper bound in Eq.~\eqref{PLOBbounda}.  
 As a consequence, we can safely remove the assumption of the ideal EPR resource $\Phi_\text{EPR}$. 

\section{Composing single-mode Gaussian channels via entanglement-assisted LOCC protocols}
We can expand our stretching property for general single-mode Gaussian channels, and address the problem how to simulate single-mode Gaussian channels via entanglement-assisted LOCC protocols in the sense of Ref.~\cite{Namiki13a}. 
Due to the unitary-equivalent decomposition of single-mode Gaussian channels \cite{Hol07,Hol08},  it is sufficient to consider a couple of specific channels described by the unitary-equivalent standard forms.  There are essentially  two standard channels (standard forms) which require entanglement for this LOCC task. This is because except for them the  standard channels belong to entanglement breaking channels, and can be simulated solely by LOCC. 

 One of the relevant standard channels is the phase-insensitive channel, which is further divided  into  the phase-insensitive lossy channel $\mathcal E_\eta^{N}$, and phase-insensitive amplification channel $\mathcal A_{\kappa} ^N$. Here, $\kappa \ge 1$ is the amplification gain and  $N \ge 0$ stands for the excess noise (See Appendix~\ref{SMGC} for detail).  Note that the pure lossy channel can be written as $  \mathcal E_\eta = \mathcal E_\eta^0 $, and the amplification channel with no excess noise  $\mathcal A_\kappa^0$ is often referred to as the quantum-limited phase-insensitive amplifier.  Essentially the same displacement-covariance property as in Eq.~\eqref{cv-t-stretch} holds for both $\mathcal E_{\eta} ^N$ and $\mathcal A_{\kappa} ^N$.  To be specific, $\mathcal A_\kappa^N $  satisfies $\mathcal A_\kappa^N  ( D(\beta )\rho   D^\dagger(\beta ) ) = 
  D(\sqrt \kappa  \beta )  \mathcal A_\kappa^N  (  \rho )  D^\dagger(\sqrt \kappa \beta )  $. 

   Using $\mathcal E_\eta^N$  instead of $\mathcal E_\eta$ in Eq.~\eqref{Sub} and following the steps toward Eq.~\eqref{Main} with  the help of   Eq.~\eqref{subloss}, 
    we obtain the effective stretching property for  general lossy channels 
   \begin{align}
 \int  \left( \mathcal L_{  D^\dagger  (\sqrt{  \eta }\xi   \beta )}   \circ \mathcal E_{\eta  }^N \circ \mathcal L_{ T_{\beta, \xi  }} (\rho ) \right )  d\! x d\! p 
=&  \mathcal E_{\eta \xi^2  }^N  (\rho ). \end{align} With the replacement $\eta \xi ^2 \to \eta $, we can write
\begin{align}
 \int  \left( \mathcal L_{  D^\dagger  (\sqrt{  \eta }  \beta )}   \circ \mathcal E_{\eta \xi^{-2}}^N \circ \mathcal L_{ T_{\beta, \xi  }} (\rho ) \right )  d\! x d\! p 
=&  \mathcal E_{\eta  }^N  (\rho ) \label{MainRLossy}, \end{align}  
where we require  $ \xi^2 > \eta $ in order that $ \mathcal E_{\eta \xi^{-2}}^N$ is a lossy channel. Essentially the same procedure for the amplification channel $\mathcal A_{\kappa   }^{ N} $with the help of Eq.~\eqref{saigo0} and some refinement yields 
\begin{align}
 \int  \left( \mathcal L_{  D^\dagger  (\sqrt{  \kappa }  \beta )}   \circ \mathcal A_{\kappa \xi^{-2}}^N \circ \mathcal L_{ T_{\beta, \xi  }} (\rho ) \right )  d\! x d\! p 
=&  \mathcal A_{\kappa   }^{\tilde N}  (\rho ) \label{MainRAMP},  \end{align}  
where \begin{align}
\tilde N = N + \kappa (1- \xi^2)/ \xi^2  . \label{tildeN}\end{align}
 Similar to the case of Eq.~\eqref{MainR}, the LHS expressions of  Eqs~\eqref{MainRLossy}~and~\eqref{MainRAMP} are regarded as entanglement-assisted LOCC protocols which precisely reproduce the channel action of the RHS terms. 
 Therefore, we can identify the entangled state $\mathcal I \otimes  \mathcal  E_{\eta \xi^{-2}}^N  (\psi_\xi )$ as a sufficient resource to simulate the channel $\mathcal E_{\eta }^N$, and 
 the entangled state $\mathcal I \otimes \mathcal   A_{\kappa \xi^{-2}}^N  (\psi_\xi )$ as a sufficient resource to simulate the channel $ \mathcal  A_{\kappa }^{\tilde N}$. 

The other relevant standard channel is the so-called additive noise channel $\mathcal E_\text{ANC}$, which becomes a phase-insensitive channel when coupled with any finite loss as is shown in Theorem 4 of Ref.~\cite{Namiki2014}. Therefore, it can be simulated as a phase-sensitive channel, practically. Since,  $\mathcal E_\text{ANC}$ is equivalent to the identity operation except it  adds a finite noise on a single quadrature variable, an exact LOCC composition  seems necessitate accomplishing the perfect teleportation as the identity operation. Formally, we can set $\kappa =1$ and $\xi \to 1 $ in Eq.~\eqref{MainRAMP} to achieve the perfect teleportation, but we must accept a finite error since we assume $\xi<1$.  

Notably, the quantum limited amplifier $\mathcal A_ \kappa ^0$ has the same constraint that the LOCC simulation comes with a finite error. This can be seen from  the fact  $\tilde N -N =0 $ holds only for $\xi =1$ in 
 Eq.~\eqref{tildeN}. 
 In sharp contrast,   the lossy channel $\mathcal E_\eta ^N$ can be  perfectly simulated without concerning the condition $\xi \to 1$,  all over the parameter space $N\ge 0$ and $\eta \in(0,1)$. In this sense,  $\mathcal E_\eta ^N$  seems exceptional, although it remains the possibility that  $\mathcal A_\kappa^0$ or  $\mathcal E_\text{ANC}$ can be perfectly simulated by a different LOCC protocol with the help of a different type of entangled resource. 

\section{Conclusion and remarks}
We have pointed out an illness of the assumption used in the proof of the tight upper bound of the ultimate communication rate over a lossy channel in Ref.~\cite{PLOB}. We have developed an effective formula of the teleportation-stretching property for lossy Gaussian channels. This enables us to  retrieve a sequence of  central inequalities appeared in the step of the proof in Ref.~\cite{PLOB} without introducing additional  assumptions. We  have also extended the resultant formula for general single-mode Gaussian channels. This extension shows how to simulate single-mode Gaussian channels via entanglement-assisted LOCC protocol in the sense of Ref.~\cite{Namiki13a}. Our result would provide a regular path to establish the definitive form of the fundamental capacities of optical quantum communication, and be widely useful to comprehend fundamental properties between  quantum entanglement and quantum channels.
It remains open to what extent  one can  generalize the present results for multi-mode Gaussian channels. 

\acknowledgments
This work was supported by the DARPA Quiness program under prime Contract No. W31P4Q-12-1-0017, the NSERC Discovery Program, and Industry Canada.

\appendix 
\section{Non-physical algebraic relations}\label{Note1}%
If we assume the CV-Bell state $\ket{\Phi_0}$ is a physical state,  
the teleportation set could be defined  as 
\begin{align}
T_\beta :=  _{A\! A^\prime}\!\! \braket{\Phi_\beta}{\Phi_0}_{A^\prime \! B} = \frac{1}{  {2\pi} } D_{B}(\beta ) \int\ket{x}_B \!\bra{x}_A dx .  \label{CVT-set}
\end{align}
The action of $T_\beta $ implies a state transfer from system $A$ to system $B$ with an addiotional displacement, namely, 
\begin{align}
T_\beta \ket{\varphi}_A=\frac{1}{ 2\pi  } D_{B}(\beta ) \ket{\varphi}_B. \end{align}
This superficially shows that  an adjustment of the local displacement $ D(\beta )$ at system $B$ accomplishes the  quantum teleportation process.

From Eqs.~\eqref{CVT-set} and \eqref{cv-t-stretch}, one can write 
\begin{align}
\mathcal L_{ D ^\dagger (\sqrt \eta \beta ) } \circ \mathcal E_\eta  \circ \mathcal L_{ T_\beta  } (\rho   )  = 
     \frac{1}{(2 \pi)^2 }\mathcal E_\eta  (  \rho )  .  
\label{cv-t-stretchR}\end{align} 
As an analogy with the finite dimension case of Eq.~\eqref{t-stretch}, one may interpret the LHS of Eq.~\eqref{cv-t-stretchR} as an LOCC protocol associated with the CV-Bell  state $\Phi_\text{0}$. However, this is no more than  an interpretation because  Eq.~\eqref{cv-t-stretchR}  over the integration of $\beta$ does not give a properly normalized output density operator.

Note that the ideal EPR state $\Phi_\text{EPR}$ in Eq.~\eqref{EPR}  has a different normalization compared with the CV-Bell state  $\Phi_\text{0}$. In fact,   its representation in the Fock basis  is given by      \begin{align}
 \ket{\Phi_{0}}   = (\sum_{n=0}^{\infty} \ketbra{n}{n}\otimes \openone)\ket{\Phi_{0}} 
= \frac{1}{\sqrt {2\pi}} \sum_{n=0}^\infty \ket{n}\!\ket{n} . 
 \end{align} 
This expression comes from Eq.~\eqref{CV-Bell} together  with the fact that  the wave function of the Fock states can be real, i.e.,~$\braket{n}{x} =\braket{x}{n}$.  

\section{Single-mode Gaussian channels} \label{SMGC}
We summarize basic description for single-mode Gaussian  channels, and useful  relations for derivation of Eqs.~\eqref{MainRLossy}~and~\eqref{MainRAMP}.
   
Any single-mode Gaussian channel can be described by a triplet of 2-by-2 matrix $(K,m,\alpha)$~\cite{Hol08}. The pair $(K, \alpha)$ and a  2-by-2 covariance matrix of  physical states $\gamma $ respectively fulfill the physical conditions 
\begin{eqnarray}
\alpha \ge \frac{i}{2}(\sigma - K^T \sigma K )  \label{PhysCon}, \  \gamma \ge \frac{i}{2} \sigma . 
\end{eqnarray}   
Here, and what follows we may use the following symbols \begin{equation}
\label{eq:S_decomposition}
\sigma : =  \left( 
\begin{array}{cc}
0  & 1 \\
-1& 0
\end{array}
\right),  \   \openone_2 : =  \left( 
\begin{array}{cc}
1  & 0 \\
0&1 
\end{array}
\right).
\end{equation}
The channel action  transforms the covariant matrix $\gamma$ and first moment $d$ of physical states  as 
\begin{eqnarray}
\gamma^\prime = K^T \gamma K + \alpha, \quad d^\prime = K^T d +m. 
\end{eqnarray}The last equation suggests $m$ is a state-independent  constant shift, and we may  set $m=0$ without notification when the constant displacement makes no physical difference.   
We can see that any composition of two Gaussian channels $\mathcal E_1$ and $\mathcal E_2$ yields another Gaussian channel $\mathcal E_{12} =  \mathcal E_2 \circ \mathcal E_1$, and the  triplet of $\mathcal E_{12}$ fulfills
\begin{eqnarray}
K_{12} &=& K_1 K_2 \nonumber, \\
m_{12} &=& K_2^T m_1 + m_2,   \nonumber \\  
\alpha_{12} &=& K_2^T \alpha _1 K_2 + \alpha_2. \label{CompoRule}
\end{eqnarray}

The lossy channel $\mathcal E _\eta^N$ is associated with  the triplet $(K , 0, \alpha )$  with
\begin{eqnarray}
K  &=& \sqrt \eta \openone_2, \ \alpha  =  \left( \frac{1-\eta}{ 2} + N \right )\openone_2 . \label{i0}
\end{eqnarray}  The pure lossy channel is the special case of  $\mathcal E _\eta^N$ with $N=0$, i.e,   $\mathcal E _\eta := \mathcal E _\eta^0$. 
Using Eqs.~\eqref{CompoRule}~and~\eqref{i0}, we can obtain the combination relation  of the pure lossy channel $ \mathcal E_\eta \circ \mathcal  E_{\eta^\prime }= \mathcal  E_{\eta \eta^\prime }$, which is  used in the derivation of Eq.~\eqref{Sub}. 
Similarly, a straightforward calculation leads to  a general formula useful to prove the relation of Eq.~\eqref{MainRLossy}: 
\begin{eqnarray}
 \mathcal E_\eta^N  \circ \mathcal  E_{\eta^\prime }^{N ^\prime}= \mathcal  E_{\eta \eta^\prime }^{N + \eta N^\prime }.\label{subloss}
 \end{eqnarray}    

The amplification channel $\mathcal A _\kappa^N$ is described by the triplet $(K , 0, \alpha )$  with
\begin{eqnarray}
K  &=& \sqrt \kappa \openone_2, \ \alpha  = \left(\frac{ \kappa -1 }{2} + N \right )\openone_2 . \label{i}
\end{eqnarray}
For  a possible  composition of the lossy channel and the amplification channel, we can show  from  Eqs.~\eqref{CompoRule},~\eqref{i0},~and~\eqref{i} that the relation 
\begin{eqnarray}  \mathcal A _\kappa^N \circ \mathcal E_{\xi^2 } =  \mathcal A _{\kappa\xi^2}^{N + \kappa (1-\xi^2)}  \label{saigo0}  \end{eqnarray} holds for  $\kappa \xi^2 \ge 1 $. A refinement of this relation yields
\begin{eqnarray}  \mathcal A _{\kappa \xi^{-2}}^N \circ \mathcal E_{\xi^2 } =  \mathcal A _{\kappa }^{N + \kappa \left( \frac{ 1-\xi^2}{\xi^2}\right)} , \label{saigo1} \end{eqnarray} where  we assume $\kappa  \ge 1 $. The relation in Eq.~\eqref{saigo0} or its refined form in Eq.~\eqref{saigo1} is essential for the derivation of Eq.~\eqref{MainRAMP}.


%
\end{document}